# Occipital and left temporal EEG correlates of phenomenal consciousness

Vitor Manuel Dinis Pereira
Language, Mind and Cognition Research Group (LanCog).
Philosophy Centre (CFUL).
Faculty of Letters, University of Lisbon
Alameda da Universidade, 1600-214 Lisboa, Portugal
vpereira1@campus.ul.pt

# Occipital and left temporal EEG correlates of phenomenal consciousness

Abstract

In the first section, Introduction, we present our experimental design.

In the second section, we characterize the grand average occipital and temporal electrical activity correlated with a contrast in access.

In the third section, we characterize the grand average occipital and temporal electrical activity correlated with a contrast in phenomenology and conclude characterizing the grand average occipital and temporal electrical activity co-occurring with unconsciousness.

Acknowledgements

My mother, Maria Dulce;

João Carneiro; Susana Lourenço;

Faculty of Psychology, University of Lisbon: Ana Marques; Inês Raposo; Inês Reis; Isabel Barahona;

Language, Mind and Cognition Research Group (LanCog), Philosophy Centre (CFUL), Faculty of Letters, University of Lisbon: Adriana Graça; João Branquinho; Ricardo Santos.

My research would not have been possible without their helps.

## Introduction

If there are occipital and temporal correlates of a stimulus about which we have no consciousness, our unconsciousness will be not only not having access but also not having phenomenal experience (Block, 2005), despite that there is electrical activity in the occipital and temporal lobes co-occurring with these stimuli.

The occipital and temporal electrical activity co-occurring with our visual experience of a stimulus will need co-occurring with consciousness, but an explanation of a contrast in access (e. g., correct and incorrect responses, namely the interval between the termination of a target and of a mask), does not explain a contrast in phenomenology (e. g., degrees of visibility, namely the mean rank within a interval of degrees of visibility) and occipital and temporal electrical activity co-occurring with a stimulus about which we have no consciousness (if any) will have to be distinguished from access and from phenomenal consciousness.

## Participants

Twenty two adults with normal vision or corrected to normal, without neurological or psychiatric history, ignoring completely the experimental purposes. Five participants were excluded dues to excessive EEG artifacts (3) or insufficient trials (2).

The experimental protocol was approved by the doctoral program in Cognitive Science, University of Lisbon.

## Apparatus and stimuli

Two types of targets: square (1.98 cm side) or diamond (for 45 ° rotation of the square).

Two types of masks: mask or pseudo-mask.

The width of the mask is 3.05 cm and its inner white portion (RGB 255-255-255) is 8 mm wider than the black (RGB 0-0-0) target stimulus.

The width of the pseudo-mask is 3.10 cm and its inner white portion is circular (2.63

cm diameter).

Despite the different sizes, the color black stands in the same area, both in mask and pseudo-mask, and its luminance is identical. This was expected to be important to make the masks produce similar ERPs when presented alone.

All stimuli are presented on a gray background (RGB 173-175-178). (Figs. 2.1-2.6.)

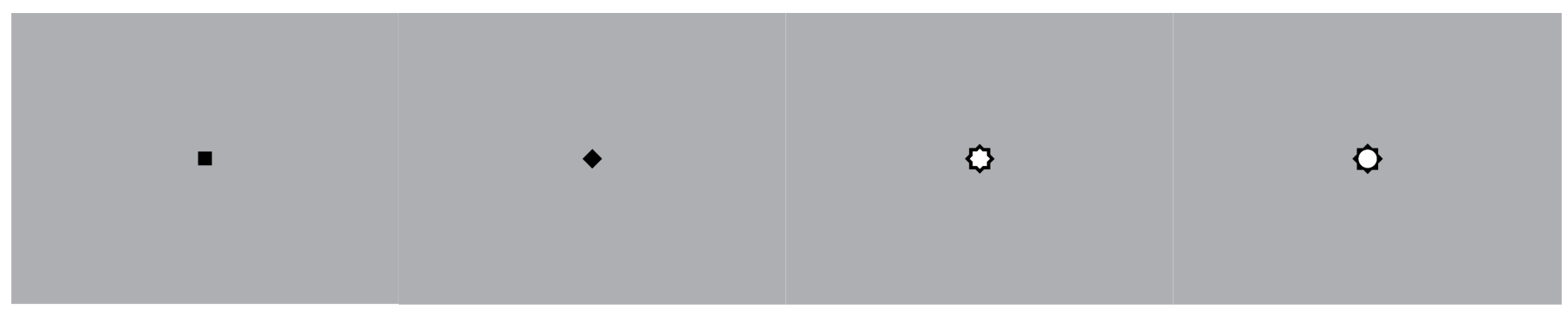

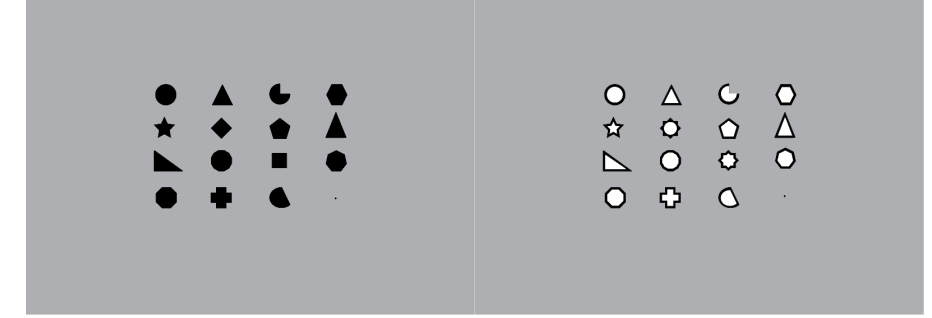

Procedure

First task: to recognize which of the targets – square or diamond – is presented.

Second task: to evaluate the visibility of targets.

The answers are given using the keyboard or the mouse.

In the first task, recognition of targets, participants respond if "they seemed to have seen something" by pressing "8" to "yes” and "9" to "no". A negative response completes the trial and starts the next. An affirmative answer conduces the subjects to a screen of sixteen stimuli to signal with the mouse what seems they have been identified. The position on the screen indicated by the participant will be recorded informatically as coordinate system <X,Y>.

In the second task (evaluation of visibility) we used a Likert scale from "1" to "5": "not visible at all" ("1" key), "barely visible" ("2" key), "visible, but obscure" (key" 3 "), "clear but not quite visible" (key "4 ") and "perfectly clear and visible" ("5" key).

Experiments were held at the Faculty of Psychology in a slightly darkened silent room. Participants were seated in a reclining chair at 81.28 centimeters distance from the 50.8 centimeters monitor.

It is expected that the running of experiment train the volunteer. The beginning of the behavioral and EEG recording is unknown to the volunteer.

The SuperLab program for Windows from Cedrus, PC - compatible, connected to a SVGA color monitor, manages the presentation of stimuli, randomizes their sequence (the trials in each block), the exposure times, the record of the response, triggers the trigger synchronize with the system acquisition of physiological signals, MP100 and EEG amplifiers, program AcqKnowledge, both of Biopac.

Statistical analysis of neuroelectric signals and behavioral data is performed using the SPSS Statistics PASW 18.

EEG recording

Electrodes of silver chloride, were placed on the scalp, according to the International System 10/20. Data were recorded with referential montage, with three channels of EEG, occipital (Oz), left temporal (T5), and right temporal (T6), referred to left mastoid.

The impedance of all electrodes was kept below 5 k Ohm. The EEG records have a duration of 1150 ms, defined as 150 ms before the stimulus (baseline) and 1000 ms after its occurrence.

If the ERP is an evoked signal (i.e. a signal super-imposed upon and independent of the ongoing noise EEG), and not a phase alignment of the ongoing signal EEG, or some combination of the two (a clear conceptual exposition of the difference is provided, for example, by Pfurtscheller and Lopes da Silva, 1999), it makes perfect sense, after visual inspection and rejection of artefacts in EEG samples for each type of sequence of stimuli, calculate the average and make the regression for the 150 ms baseline, measuring the amplitude and latency of the waveforms thus collected.

Experiment I

Eight volunteers (aged 18–46 years, M= 22.50, SD=9.562, 7 females).

The target and the masks will be presented for 17 ms. The mask (or pseudo-mask) appears 1 ms after the presentation of the target (inter-stimulus interval, ISI, the interval between the termination of the target and the onset of the mask). These ISI (1 ms) correspond to 18 ms stimulus-onset asynchrony (SOA, the interval between the onset of the target and the mask) (rounded values). Answers were signaled by mouse on a screen of sixteen stimuli, among which are the mask and pseudo-mask. Note that the subject not performed a forced-choice task, for example, reading any question either ''Diamond or Square?'' or ''Square or Diamond?'', even if counterbalanced across participants (contrast with, for example, Lau and Passingham, 2006).

In the second trial, masks were presented for 17 ms, and answers were signaled by mouse on a screen of sixteen stimuli, among which are the mask and pseudo-mask.

Second block. Trial: targets will be presented for 17 ms, and answers were signaled by mouse on screen of sixteen stimuli, among which are the targets. (Figs.

2.2-2.4.)

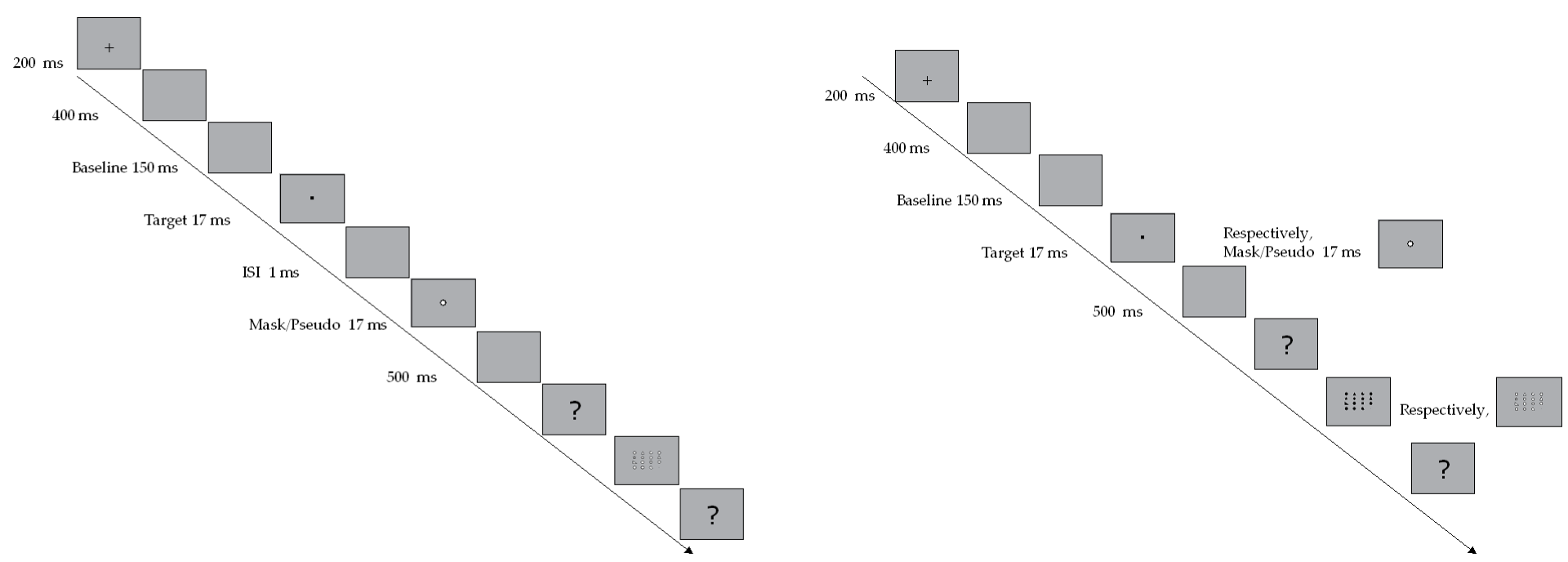

Experiment II

Nine volunteers (aged 20–26 years, M= 21.22, SD=2.224, 5 females).

The target will be presented for 17 ms (like experiment I), but the mask (or pseudo-mask) will be presented for 167 ms (unlike experiment I). Note that in all experiments, targets are always shown for 17 ms and is never replaced, for example, by a blank screen with the same duration of 17 ms (contrast, for example, with Del Cul et al. 2007).

Unlike experiment I, the target is intercalated between two presentations of the mask/pseudo-mask (each, 167 ms): one earlier, paracontrast; the other after target, metacontrast.

The mask (or pseudo-mask) appears 0 ms before (forward masking) and 1 ms after (backward masking) the presentation of the target (inter-stimulus interval, ISI, the interval between the termination of the target and the onset of the mask). These ISI (1 ms) correspond to 18 ms stimulus-onset asynchrony (SOA, the interval between the onset of the target and the mask) and to 168 ms stimulus-termination asynchrony (STA, the interval between the termination of the target and of the mask) (rounded values). Unlike experiment I, the answer were signaled by mouse on screen of sixteen stimuli, among which are the targets.

In the second trial, like experiment I: masks were presented for 17 ms, and answers were signaled by mouse on a screen of sixteen stimuli, among which are the mask and pseudo-mask.

Second block, like experiment I. Trial: targets will be presented for 17 ms, and answers were signaled by mouse on screen of sixteen stimuli, among which are the targets. (Figs. 2.3-2.4.)

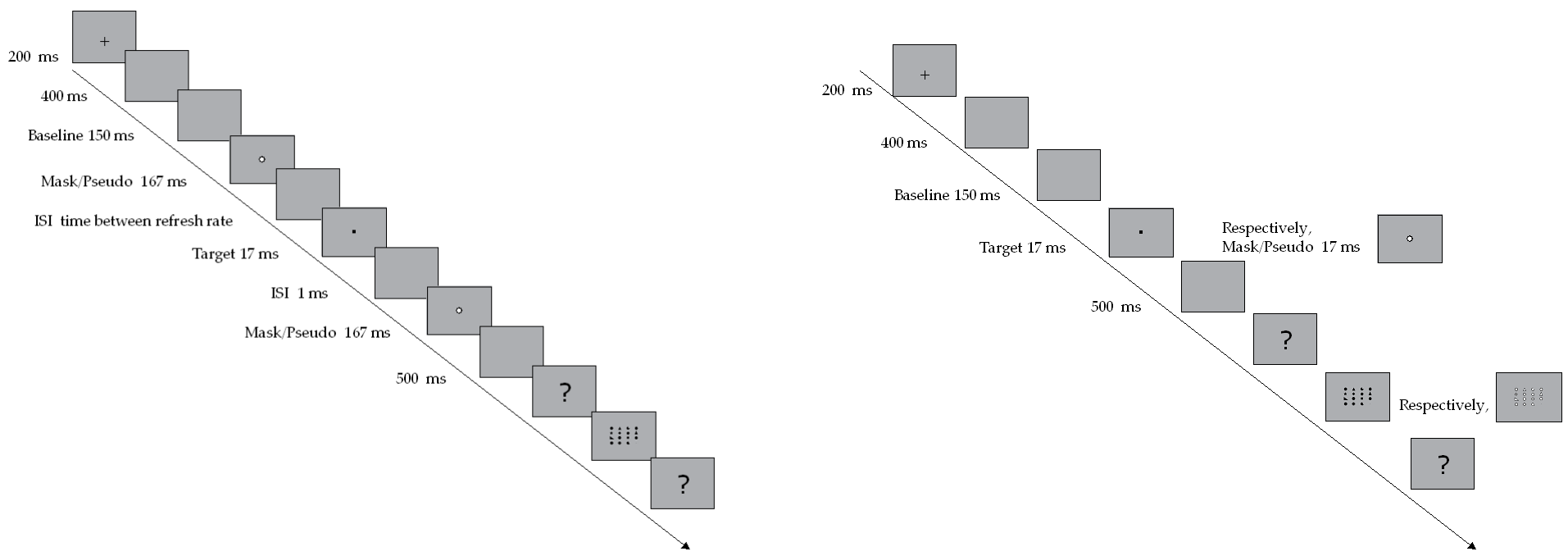

The grand average occipital and temporal electrical activity correlated with a contrast in access

ERPs data

The Oz and T5 maximum positive amplitude 300-800 ms of the ERPs for experiment II is greater compared to that of experiment I for combined target-mask presentations than for isolated presentations.

The repeated measures ANOVA (figs. 2.5-2.8) with experiment I/ experiment II

as a between-subjects factors and electrophysiological maximum positive amplitude 300-800 ms related to events combined target-mask/ isolated presentations as a within-subject factors gave the following significant (Greenhouse/Geisser Correction for violations of the sphericity) results for Oz [F(2.557,38.349)= 3.348, p < 0.035] and T5 [F(3.700, 55.507)= 4.066 p < 0.007], but no significant results for T6 [F(1.223, 18.342)= 0.774 p < 0.416].

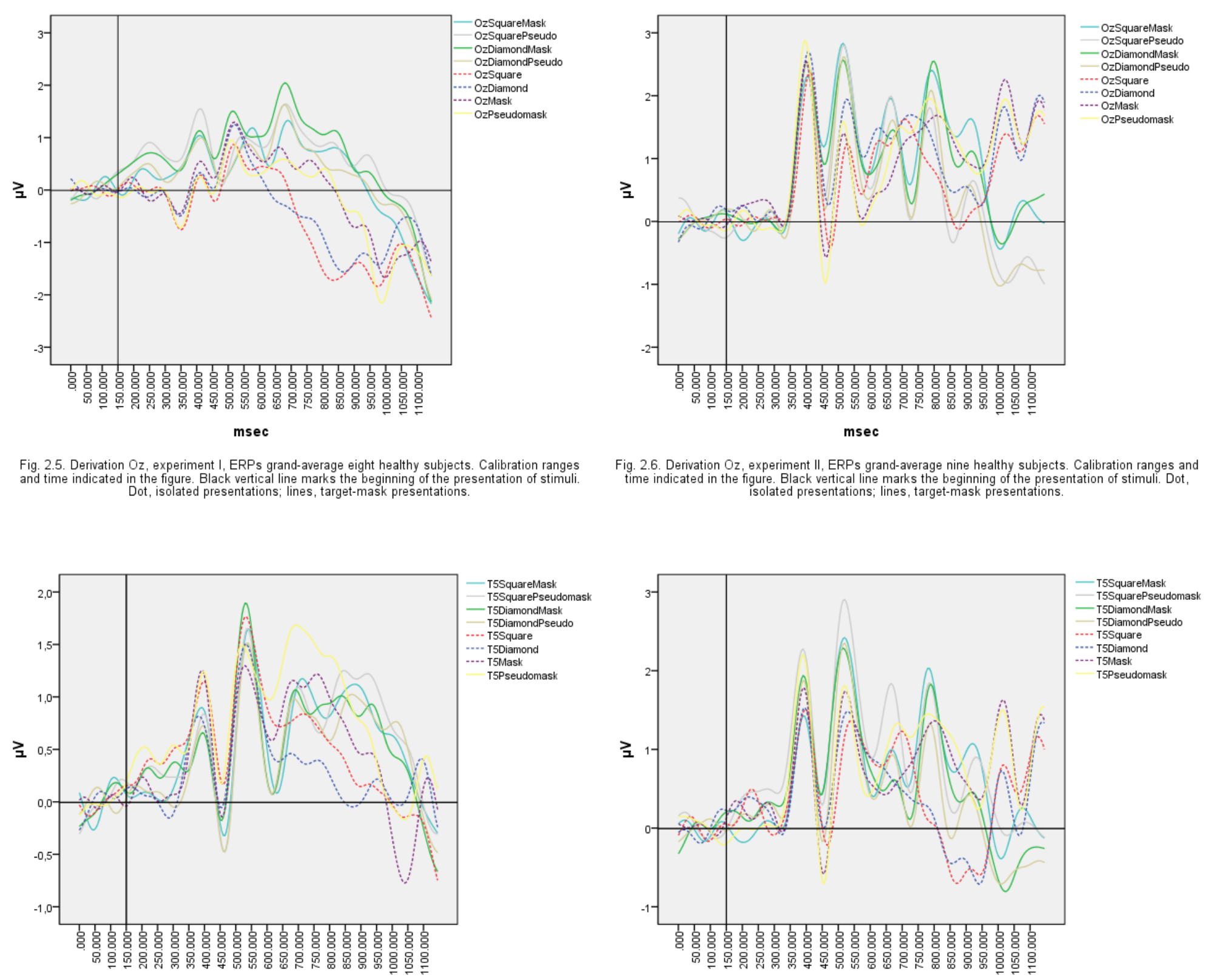

Fig. 2.5. Derivation Oz, experiment I, ERPs grand-average eight healthy subjects. Calibration ranges and time indicated in the figure. Black vertical line marks the beginning of the presentation of stimuli. Dot, isolated presentations; lines, target-mask presentations.

Fig. 2.6. Derivation Oz, experiment II, ERPs grand-average nine healthy subjects. Calibration ranges and time indicated in the figure. Black vertical line marks the beginning of the presentation of stimuli. Dot, isolated presentations; lines, target-mask presentations.

Fig. 2.7. Derivation T5, experiment I, ERPs grand-average eight healthy subjects. Calibration ranges and time indicated in the figure. Black vertical line marks the beginning of the presentation of stimuli. Dot, isolated presentations; lines, target-mask presentations.

Fig. 2.8. Derivation T5, experiment II, ERPs grand-average nine healthy subjects. Calibration ranges and time indicated in the figure. Black vertical line marks the beginning of the presentation of stimuli. Dot, isolated presentations; lines, target-mask presentations.

Computation of SSE0, deltaBIC, the Bayes factor, and the posterior probabilities for the null and alternative hypotheses from input consisting of n (number of independent observations), k1 – k0 (the difference between the two models with respect to number of free parameters), sum of squares for the effect of interest, and sum of squares for the error term associated with the effect of interest (SSE1)[1], as implemented in Excel by Masson (2011), from the repeated measures ANOVA for the ERPs collected in our experiments I and II, for Oz is n = 119, df_effect = 2.557, SSeffect = 27.716, SSerror = 124.195, SSE1 = 124.1954, SSE0 = 151.9119, deltaBIC = -11.7536, BF01 = 0.002804, p(H0|D) = 0.002796, p(H1|D) = 0.997204.

For T5 is n = 119, df_effect = 3.700, SSeffect = 12.735, SSerror = 46.98398, SSE1 = 124.1954, SSE0 = 59.71874, deltaBIC = -10.8557, BF01 = 0.004392, p(H0|D) = 0.004373, p(H1|D) = 0.995627.

For T6 is n = 119, df_effect = 1.223, SSeffect = 36.397, SSerror = 705.457, SSE1 = 705.4575, SSE0 = 741.8546, deltaBIC = -0.1427, BF01 = 0.931137, p(H0|D) = 0.48217, p(H1|D) = 0.51783.

[1] Note that the term SSE1/SSE0 is just the complement of partial eta-squared ($\eta p^2$ ), a measure of effect size corresponding to the proportion of variability accountedfor by the independent variable (i.e., SSE1/SSE0= 1- $\eta p^2$ ).

Behavioral data

The subjects gave more correct responses than incorrect responses (first task, see Procedure) in target-mask experiment I (the answer is signaled by mouse on a screen of sixteen stimuli among which are the mask and pseudo-mask: correct responses 94.47%, incorrect 5.52%) (fig. 2.9) than in target-mask experiment II (the answer is signaled by mouse on the screen of sixteen stimuli among which are the targets: correct responses 33.67%, incorrect 66.32%) (fig. 2.10).

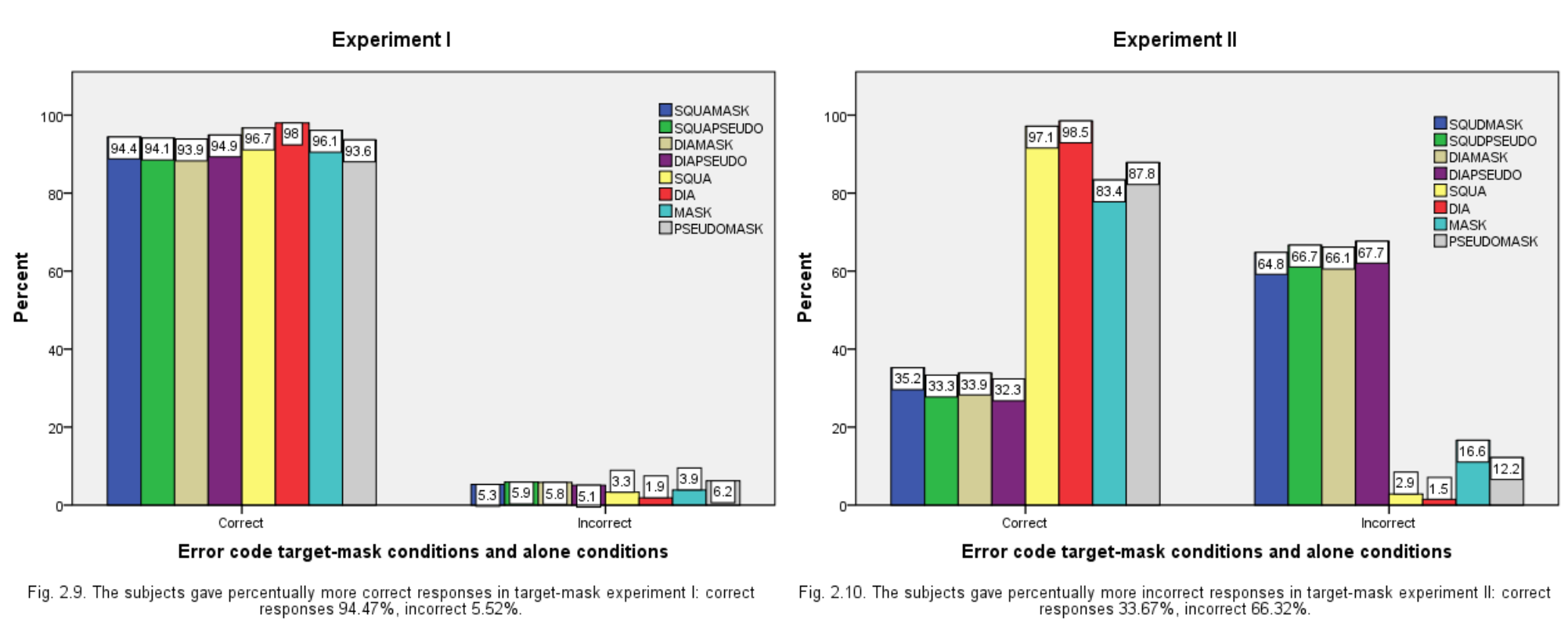

Fig. 2.9. The subjects gave percentually more correct responses in target-mask experiment I: correct responses 94.47%, incorrect 5.52%.

Fig. 2.10. The subjects gave percentually more incorrect responses in target-mask experiment II: correct responses 33.67%, incorrect 66.32%.

The target-mask presentations hinders the task of stimuli recognition in target-mask experiment II from 5.52% (when, at experiment I, the correct discrimination would be mask or pseudo-mask, dependent on if had been shown a mask or a pseudo-mask) to 66.32% of incorrect responses (when, at experiment II, the correct discrimination would be square or diamond, dependent on if had been shown a square or a diamond).

Given that in target-mask experiment II, the target (presented for 17 ms) is paracontrast and metacontrast by the mask/pseudo-mask (presented for 167 ms), it is the 168 ms (rounded values) interval between the termination of the target and of the mask that explain why (at least, with I and II experimental design) the subject response is very often "none of the fifteen stimuli presented" (the dot in the right bottom of answers screen) (fig. 2.11), none of the targets, despite that in all experiments, targets are always shown.

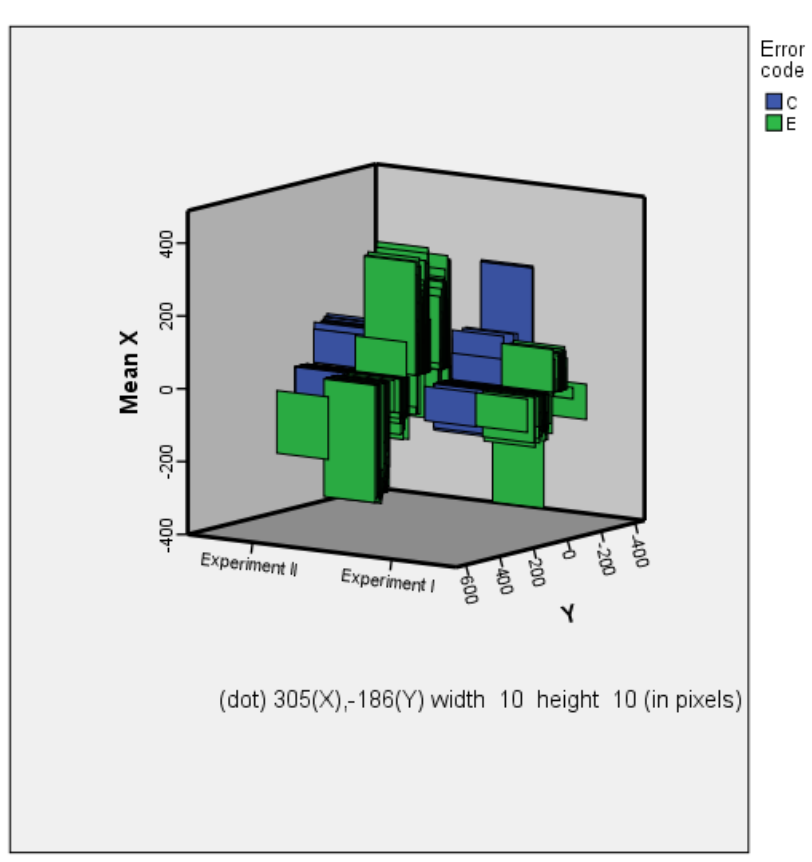

In other words, information about if the target is paracontrast and metacontrast by the mask/pseudo-mask, 18 ms SOA and 168 ms STA helped us improve our prediction of the stimulus discrimination by 43,9% in experiment II (the lambda asymmetric measure of association is .439) and information about if the target is metacontrast by the mask/pseudo-mask and 18 ms SOA helped us improve our prediction of the stimulus discrimination by 0 % in experiment I (the lambda asymmetric measure of association is .000). That is, the proportion of relative error in predicting stimulus discrimination that can be eliminated by knowledge of the way stimulus are display is .439 for experiment II and is .000 for experiment I.

And information about stimulus discrimination helped us improve our prediction that the target is paracontrast and metacontrast by the mask/pseudo-mask, 18ms SOA and 168ms STA by 9.3 % in experiment II (the lambda asymmetric measure of association is .093) and information about stimulus discrimination helped us improve our prediction that the target is metacontrast by the mask/pseudo-mask, 18ms SOA by 0.5 % in experiment I (the lambda asymmetric measure of association is .005). That is, the proportion of relative error in predicting the way stimulus are display that can be eliminated by knowledge of stimulus discrimination is .093 for experiment II and is .005 for experiment I.

Thus the contrast in Oz and T5 between more high positive amplitude 300-800 ms in combined stimuli presentations for experiment II than for experiment I correlate with the statistically significant contrast in stimuli discrimination between more incorrect responses in experiment II [$\chi(7) = 2370.368$, $p = .000$, 0 cells (.0%) have expected count less than 5. The minimum expected count is 293.80] than in experiment I [$\chi(7) = 27.029$, $p = .006$, 0 cells (.0%) have expected count less than 5. The minimum expected count is 29.83] and so correlate with contrast in access (the grand average ERPs correlated with correctly identify the stimulus would be the correlate of the access). (The Chi square test χ2 is carried out on the actual numbers of correct and incorrect responses, not on percentages, proportions, means or other derived statistics of correct and incorrect responses.) (Figs. 2.12-2.13.)

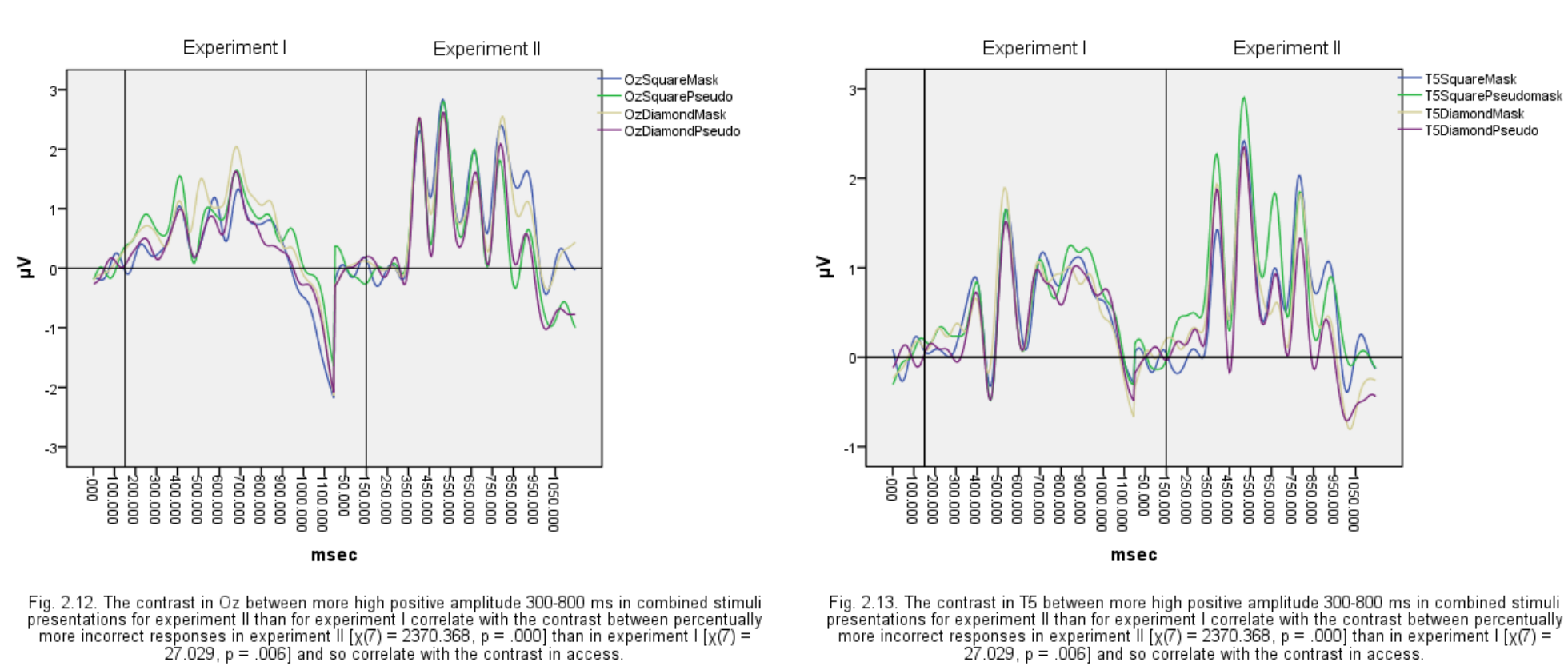

Fig. 2.12. The contrast in Oz between more high positive amplitude 300-800 ms in combined stimuli presentations for experiment II than for experiment I correlate with the contrast between percentually more incorrect responses in experiment II [χ(7) = 2370.368, p = .000] than in experiment I [χ(7) = 27.029, p = .006] and so correlate with the contrast in access.

Fig. 2.13. The contrast in T5 between more high positive amplitude 300-800 ms in combined stimuli presentations for experiment II than for experiment I correlate with the contrast between percentually more incorrect responses in experiment II [χ(7) = 2370.368, p = .000] than in experiment I [χ(7) = 27.029, p = .006] and so correlate with the contrast in access.

The grand average Oz ERPs positive amplitude 300-800 ms for combined presentations is 3.678 µV for experiement II and 2.5145 µV for experiment I.

And the grand average T5 ERPs positive amplitude 300-800 ms for combined presentations is 3.139 µV for experiement II and 2.582 µV for experiment I.

## The grand average occipital and temporal electrical activity correlated with a contrast in phenomenology

Notwithstanding, in target-mask presentations of the experiment II, the degree of visibility (second task, see Procedure) range from 19.27% "clear but not quite visible" (key "4") to 62.22% "perfectly clear and visible" (key "5"). (Fig. 2.14.)

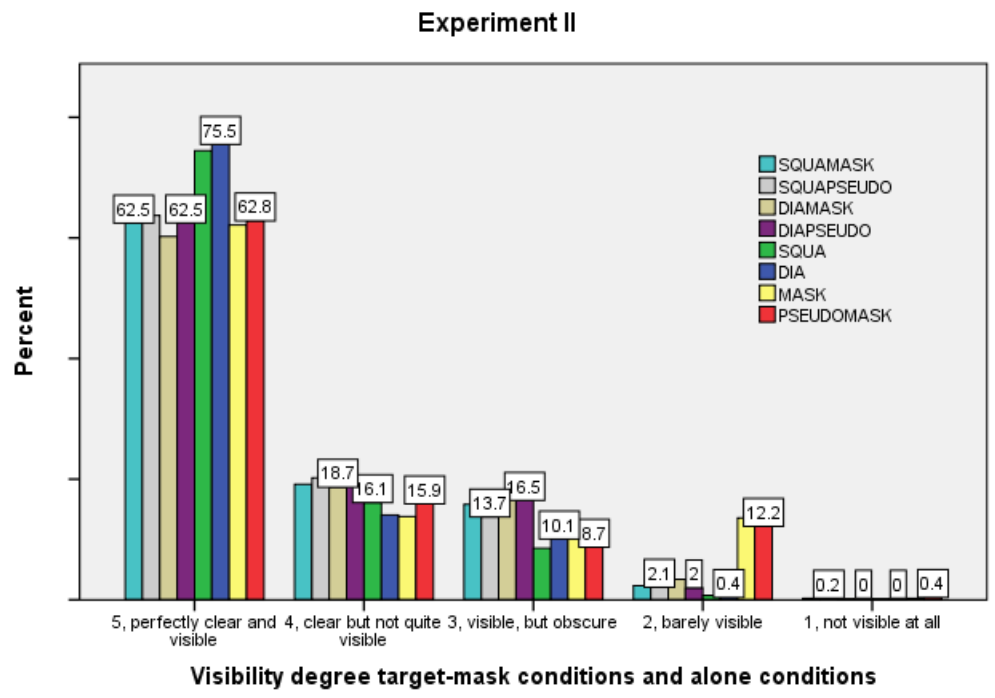

Fig. 2.14. In target-mask presentations of the experiment II, the visibilty range from 19.27% "clear but not quite visible" (key "4") to 62.22% "perfectly clear and visible" (key "5").

However, in target-mask presentations of the experiment I, range from 31.05% "clear but not quite visible" (key "4") to 32.52% "perfectly clear and visible" (key "5").(Fig. 2.15.)

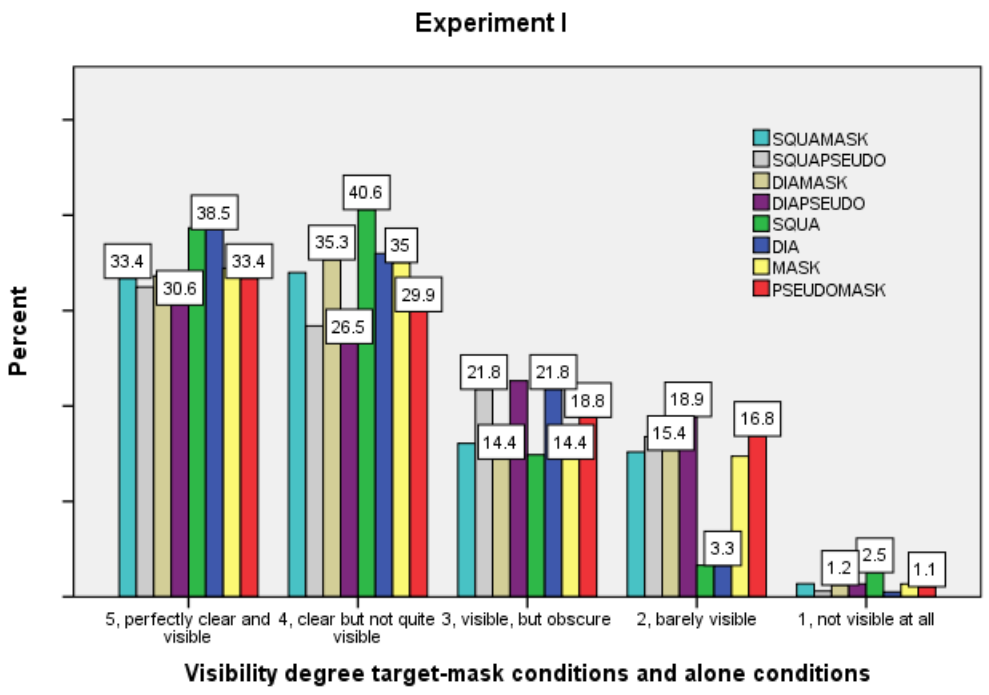

Fig. 2.15. In target-mask presentations of the experiment I, the visibilty range from 31.05% "clear but not quite visible" (key "4") to 32.52% "perfectly clear and visible" (key "5").

Average reaction time appears not to be the better explanation: experiment I [RT for correct responses, 1055 ms; RT for incorrect responses, 1129 ms]; experiment II [RT for correct responses, 1338 ms; RT for incorrect responses, 1715 ms]. (Fig. 2.16.)

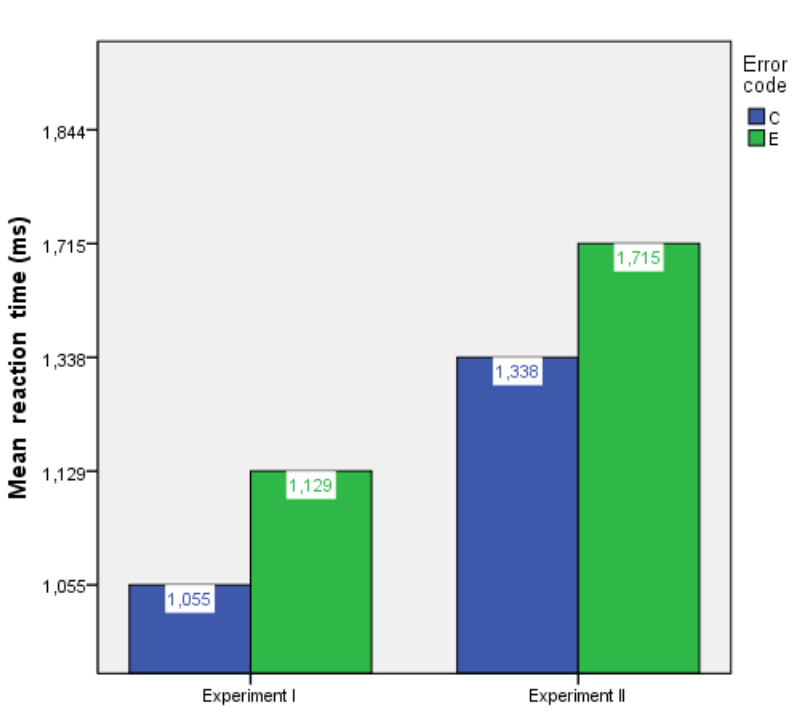

Fig. 2.16. Mean stimuli discrimination reactions times not appears to be the better place to look: experiment I [RT for correct responses, 1055 ms; RT for incorrect responses, 1129 ms]; experiment II [RT for correct responses, 1338 ms; RT for incorrect responses, 1715 ms].

One hypothesis should be that the subjects take on average less time to respond incorrectly in experiment II than in experiment I. However the subjects take on average longer to respond incorrectly in experiment II than in experiment I.

Even if we look to keys “4” and “5” mean reactions times, average reaction time not appears to be the better explanation. (Fig. 2.17.)

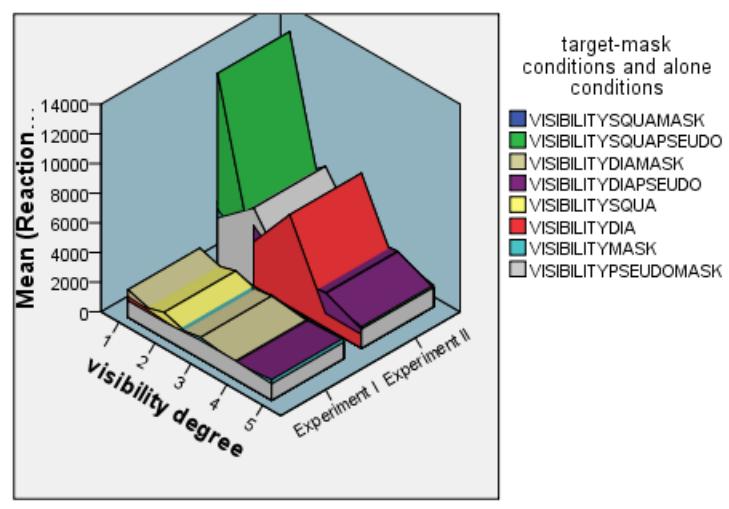

Other hypothesis should be that the subjects take on average less time to press the keys “4” or “5” in experiment II than in experiment I. However the subjects take on average longer to press the keys “4” or “5” in experiment II than in experiment I.

Meanwhile the difference between the two experiments in mean rank within the interval of degrees of visibility "4" and "5" is statistically significant either if the targets combined with masks or with pseudo-masks were incorrectly or correctly identified [H(1) = 290.908, p = 0.000, with a mean rank of 1848.19 for experiment I, and a mean rank of 2393.18 for experiment II (Kruskal-Wallis Test)] or if the targets combined with masks or pseudo-masks were correctly identified [H(1) = 294.905, p = 0.000, with a mean rank of 1825.70 for experiment I, and mean rank of 2372.96 for experiment II (Kruskal-Wallis Test)]. If the targets combined with masks or pseudo-masks were incorrectly identified the difference between the two experiments in mean rank within the interval of degrees of visibility "4" and "5" is statistically significant [H(1) = 291.513, p = 0.000, with a mean rank of 1837.76 for experiment I, and mean rank of 2382.65 for experiment II (Kruskal-Wallis Test)].

Note that while in experiment II the high degrees of visibility is not assigned to the target that were asked the participants to identify in combined target-mask presentations (but to the mask/pseudo-mask they saw), in experiment I the high degrees of visibility is assigned to the mask/pseudo-mask that were asked the participants to identify in combined target-mask presentations.

In both experiments I and II the high degrees of visibility is assigned to the mask/pseudo-mask they saw in combined target-mask presentations. Incorrectly, in experiment II combined target-mask presentations: at experiment II combined target-mask presentations, the correct discrimination would be square or diamond, dependent on if had been shown a square or a diamond. Correctly, in experiment I combined target-mask presentations: at experiment I combined target-mask presentations, the correct discrimination would be mask or pseudo-mask, dependent on if had been shown a mask or a pseudo-mask.

If the difference between the two experiments in mean rank within the interval of degrees of visibility "4" and "5" remains statistically significant between the two experiments in targets isolated presentations correctly identified, the interval between the termination of the target and of the mask does not explain the contrast in mean rank within the interval of degrees of visibility "4" and "5" (the targets correctly identified its is presented isolatedly and so without any mask).

There must be another explanation for the contrast between the two experiments in targets isolated presentations correctly identified in the variability of the mean rank within the interval of degrees of visibility "4" and "5".

The procedure for the classification of the Kruskal-Wallis Test is as follows: the lowest grade, in this case "4", receives the minor rating, in the case of "4", receives classification "1"and the next grade, in this case "5", receives classification "2".

The second trial of the first block and the second block - in which the stimuli are presented in isolation - are the same in both experiments I and II, thus it is expected that the discrimination of stimulus not statistically significant contrast in correct and incorrect responses between the two experiments.

In isolated presentations of targets, there is no statistically significant contrast in stimuli discrimination between the two experiment II [$\chi(1) = 3.492$, $p = .062$, 0 cells (.0%) have expected count less than 5. The minimum expected count is 17.47] and experiment I [$\chi(1) = 3.160$, $p = .075$, 0 cells (.0%) have expected count less than 5. The minimum expected count is 20.38], the contrast in stimuli discrimination between more correct responses in experiment II (square correct 97.1%, incorrect 2.9 %; diamond correct 98.5%, incorrect 1.5 %) than in experiment I (square correct 96.7%, incorrect 3.3 %; diamond correct 98.1 %, incorrect 1.9 %) is not statistically significant, and so, there are not a difference statistically significant between correct responses in experiment II than in experiment I (the block in which the targets are presented in isolation is the same second in both experiments I and II). The access not significantly change between experiment II and experiment I, the access remains the same.

However, in mask or pseudo-mask isolated presentations, there is a statistically significant contrast in stimuli discrimination in experiment II [$\chi(1) = 6.256$, $p = .012$, 0 cells (.0%) have expected count less than 5. The minimum expected count is 112.36], contrary to a no statistically significant contrast in stimuli discrimination in experiment I [$\chi(1) = 3.682$, $p = .055$, 0 cells (.0%) have expected count less than 5. The minimum expected count is 33.35].

In mask or pseudo-mask isolated presentations, the incorrect responses are significantly greater in experiment II (mask incorrect 16.6 %, pseudo-mask incorrect 12.2 %) than in experiment I (mask incorrect 3.9%, pseudo-mask incorrect 6.2 %) and the correct responses are not significantly greater in experiment I (mask correct 96.1%, pseudo-mask correct 93.6%) than in experiment II (mask correct 83.4%, pseudo-mask correct 87.8%).

Arguably, because the second trial is randomized in the first block with the first trial where the correct identification of stimuli are targets in combined presentations in experiment II (dependent on if had been shown a square or a diamond) and masks in combined presentations in experiment I (dependent on if had been shown a mask or a pseudo-mask).

The difference between the two experiments in mean rank within the interval of degrees of visibility "4" and "5" remains statistically significant between the two experiments in targets isolated presentations correctly identified [H(1) = 336.045, p = 0.000), with a mean rank of 1081.05 for experiment I, and a mean rank of 1522.38 for experiment II (Kruskal-Wallis Test)].

The interval between the termination of the target and of the mask does not explain the contrast in mean rank within the interval of degrees of visibility "4" and "5". The better explanation is the phenomenology. If, as it is, the access is the same (fig. 2.18) and if, as there are, there are a statistically significant difference in mean rank within the interval of degrees of visibility "4" and "5", the better explanation is a difference in phenomenology.

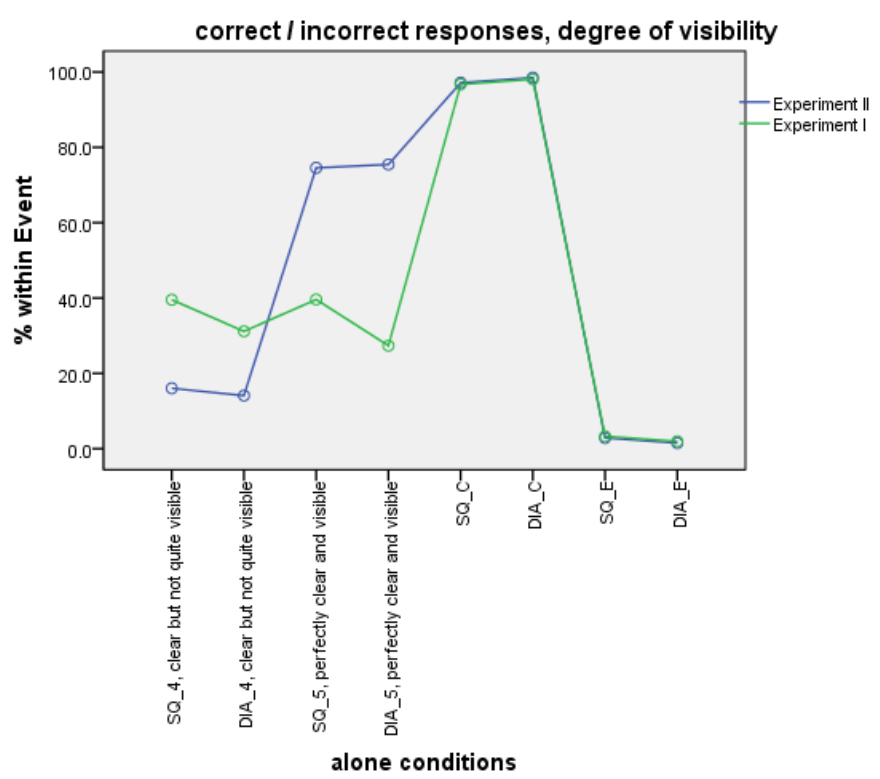

Thus the contrast between more high positive amplitude 300-800 ms in targets isolated presentations for experiment II than for experiment I (Oz, square/diamond and T5, diamond) and in T5 between more high positive amplitude 300-800 ms in square targets isolated presentations for experiment I than for experiment II (targets, because there are more incorrect responses in mask or pseudo-mask isolated presentation for II than I) correlate with the contrast between high mean rank within the interval of degrees of visibility "4" and "5" in targets isolated presentations for experiment II than for experiment I and so correlate with the contrast in phenomenology (in isolated presentations of targets, there is no statistically significant contrast in correct responses between the two experiments). (Figs. 2.19-2.22.)

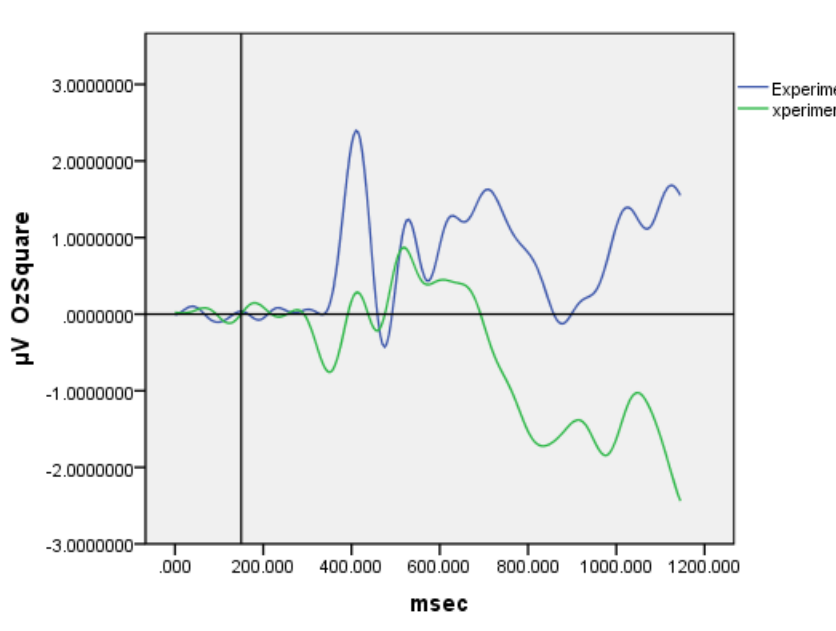

Fig. 2.19.The contrast in Oz between more high positive amplitude 300-800 ms in square isolated presentations for experiment II than for experiment I correlate with the contrast between high mean rank degrees of visibility "4" and "5" in square isolated presentations for experiment II than for experiment I and so correlate with the contrast in phenomenology

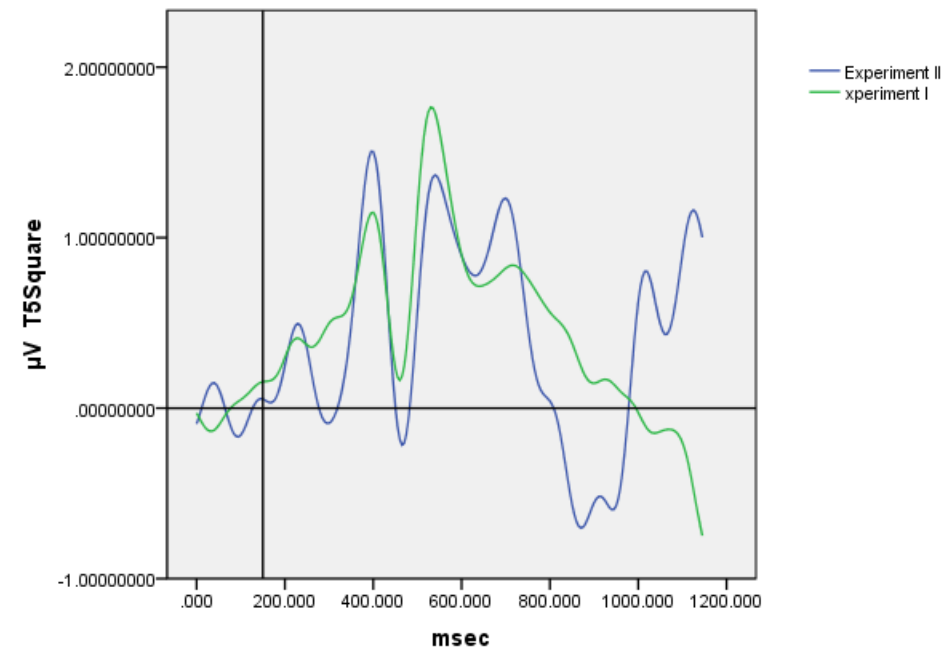

Fig. 2.20. The contrast in T5 between more high positive amplitude 300-800 ms in square isolated presentations for experiment I than for experiment II correlate with the contrast between high mean rank degrees of visibility "4" and "5" in square isolated presentations for experiment II than for experiment I and so correlate with the contrast in phenomenology

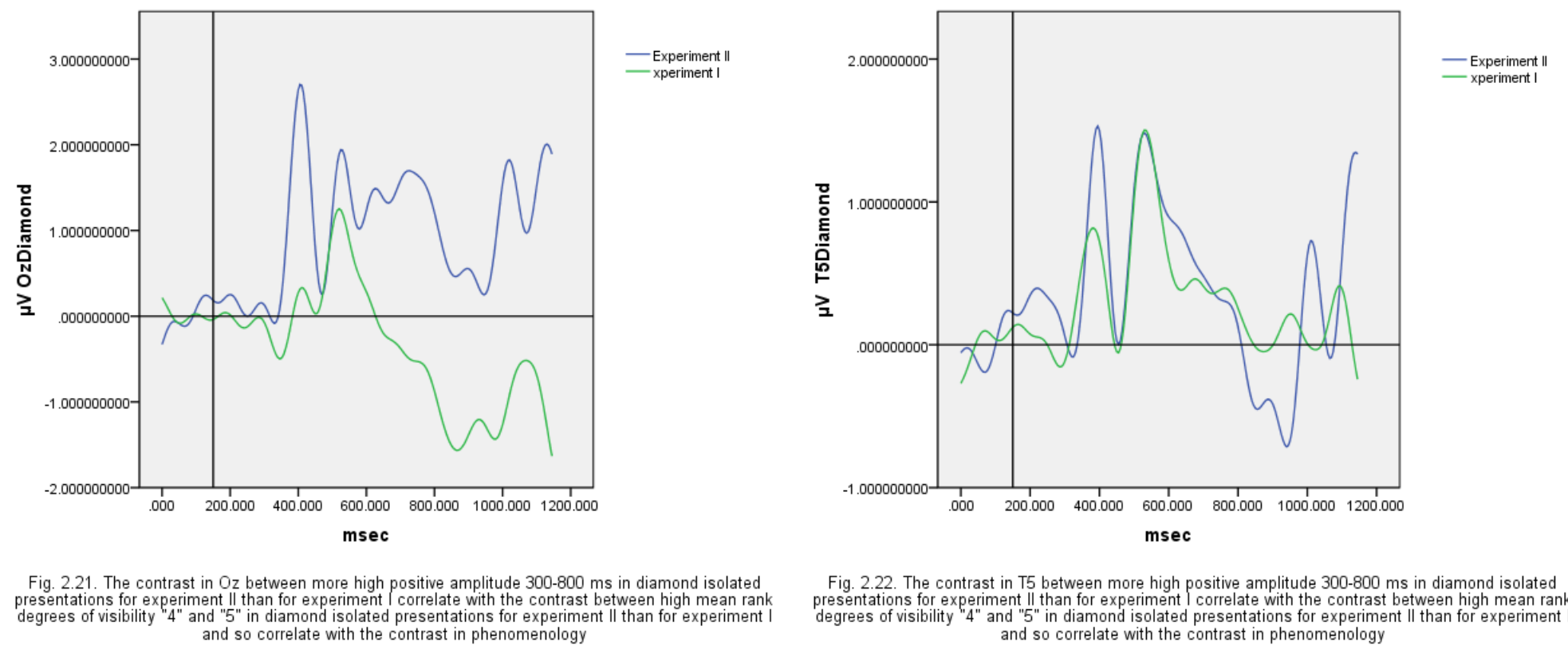

Fig. 2.21. The contrast in Oz between more high positive amplitude 300-800 ms in diamond isolated presentations for experiment II than for experiment I correlate with the contrast between high mean rank degrees of visibility "4" and "5" in diamond isolated presentations for experiment II than for experiment I and so correlate with the contrast in phenomenology

Fig. 2.22. The contrast in T5 between more high positive amplitude 300-800 ms in diamond isolated presentations for experiment II than for experiment I correlate with the contrast between high mean rank degrees of visibility "4" and "5" in diamond isolated presentations for experiment II than for experiment I and so correlate with the contrast in phenomenology

## The grand average occipital and temporal electrical activity co-occurring with unconsciousness

Given that the subject response, in target-mask experiment II, is very often "none of the fifteen stimuli presented", they have no access to the targets and given that they assigned a high degree of visibility to the mask/pseudo-mask they saw, they have no visual experience of the targets, the contrast in Oz and T5 between more high positive amplitude 300-800 ms in combined stimuli presentations for experiment II than for experiment I is not a contrast between access and phenomenal consciousness of the targets, it is a contrast in access, namely if the mask/pseudo-mask experiment II EEG signal is subtracted to the target-mask experiment II EEG signal, the result is (at least arguably) the occipital and left temporal electrical activity co-occurring with the targets about which we have no consciousness. (Figs. 2.23-2.24.)

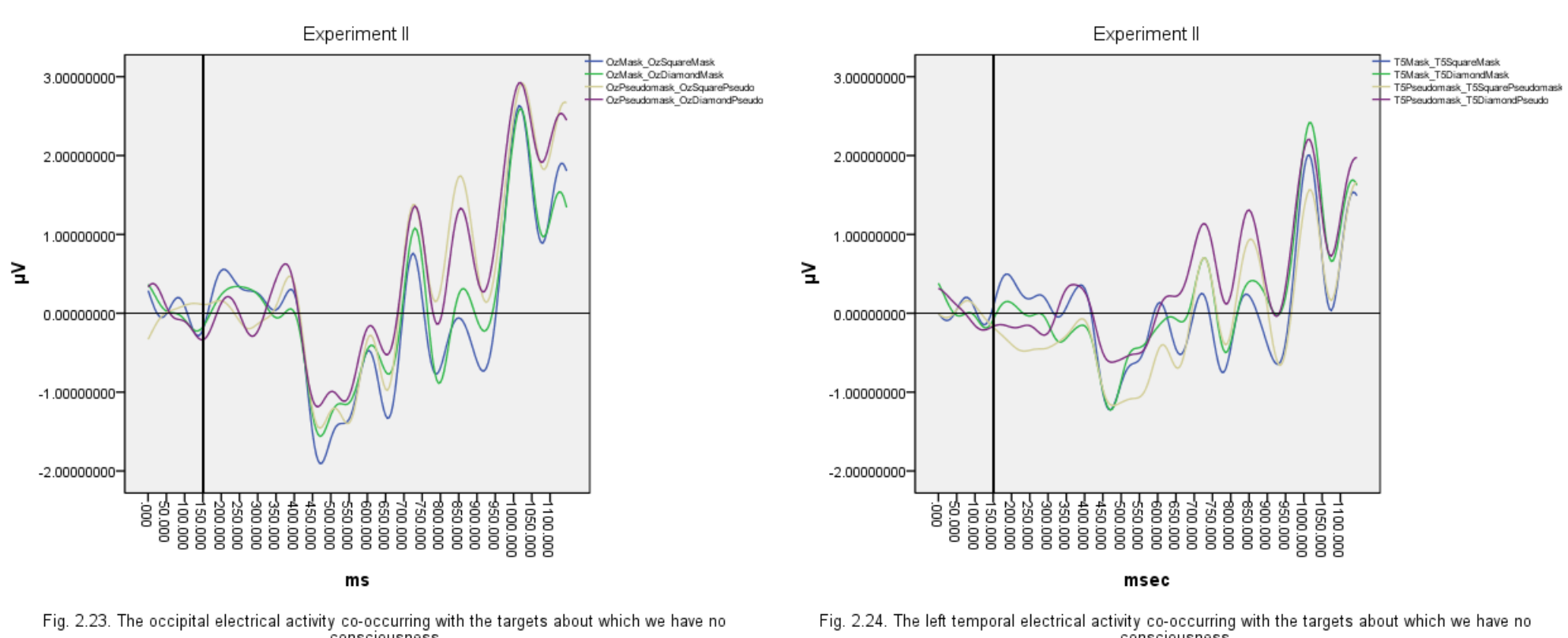

Fig. 2.23. The occipital electrical activity co-occurring with the targets about which we have no consciousness

Fig. 2.24. The left temporal electrical activity co-occurring with the targets about which we have no consciousness

# References

Block, N. (2005). Two neural correlates of consciousness. *Trends in Cognitive Sciences*, 9, 46-52.

Del Cul, A., Baillet, S., Dehaene, S. (2007). Brain dynamics underlying the nonlinear threshold for access to consciousness. *PLoS Biol*, 5, 2408-2423. (Late positivity, LP, 500-600 ms.)

Lau, H. C. & Passingham, R. E. (2006). Relative blindsight in normal observers and the neural correlate of visual consciousness. *Proceedings of the National Academy of Sciences of the United States of America*, 103, 18763–18768.

Masson, M. E. J. (2011). A Tutorial on a Practical Bayesian Alternative to Null–Hypothesis Significance Testing. *Behavior Reseach Methods,* 43, 679–690.

Pfurtscheller, G. & Lopes da Silva, F.H. (1999). Event-related EEG/MEG synchronization and desynchronization: basic principles. *Clinical Neurophysiology*, 110, 1842–1857.